\documentstyle[pra,aps,epsfig]{revtex}
\begin{document}
\draft
\twocolumn
\wideabs{
\title{Evidence for Adiabatic Magnetization of  cold  Dy{\bf ${\rm _N}$} Clusters}
\author{S. Pokrant}
\address{Laboratoire de Magn\'etisme Louis Ne\'el, B.P. 166, F-38042 Grenoble Cedex 9}
\date{\today}
\maketitle
\begin{abstract}
Magnetic properties of ${\rm Dy_N}$ clusters in a molecular beam
generated with a liquid helium cooled nozzle are investigated by
Stern-Gerlach experiments. The cluster magnetizations $\mu_z$ are
measured as a function of magnetic field  
($B=0-1.6$T) and  cluster size ($17 \le {\rm N} \le 55$).
The most important  observation is the saturation of the magnetization 
 $\mu_z(B)$ at large field strengths. The magnetization
  approaches  saturation following the power
law $|\mu_z-\mu_0|\sim 1/\sqrt{B}$, where $\mu_0$ denotes the magnetic moment. 
This gives evidence for  adiabatic magnetization.
\end{abstract}
}
\narrowtext
% body of paper here
%\section{Introduction}
The first Stern-Gerlach experiments
\cite{Ste22} on metal clusters
 have been performed on ${\rm Fe_N}$ with N=15-650 \cite{deH90}.
  Surprisingly,  in these experiments the ${\rm Fe_N}$
clusters were deflected only in the direction of the increasing
magnetic field, in contrast to the well known deflection patterns of
atoms and small molecules \cite{Ged89,Kue88}. Since then the question of the
 magnetization process of small isolated clusters  has  been widely
discussed in literature.
Khanna et al. suggested that the spins of the metal atoms
in the clusters are coupled superparamagnetically resulting in one large
spin \cite{Kan91}.
If the thermal energy is substantially
larger than the magnetic anisotropy  energy,
the classical Langevin model can be applied to explain the magnetization
of the clusters as a function of  field $B$, cluster size $N$ and internal
cluster temperature $T_c$. According to this model the freely fluctuating spin
 relaxes under isothermal conditions, since
the cluster vibrations seem to provide a
  heat bath which is sufficiently large
  to maintain a constant temperature during the
 magnetization   process, i.e. when
the clusters enter the magnetic field.  Most of the
experiments performed to investigate the magnetic properties
 of small metal particles were carried out at high temperatures.
 The results can be understood within an isothermal magnetization model
 using $\mu_z={\mathcal L}(\mu_0B/k_BT_c) \mu_0$, where $k_B$ denotes the
 Boltzmann constant and $\mathcal{L}$ the Langevin function.
 Assuming  $T_c\approx T_n$, with $T_n$ denoting the nozzle
 temperature, the approximation $\mu_z=\mu_0 (\mu_0B/3k_BT_c)$ in
 the low field limit $(\mu_0B\ll k_BT_c)$ was applied to fit the magnetization
 curves. 
  The obtained values
 for the magnetic moments per atom $\mu_0/N$ of large ${\rm Co_N}$ \cite{Bil94},
 ${\rm Fe_N}$ \cite{Bil93} and ${\rm Ni_N}$ \cite{Blo96} clusters
 with N$\ge 2500$ were consistent with macroscopic
 magnetic properties of Fe, Co and Ni. However, the  saturation of the
  magnetization predicted for the strong field
limit $(\mu_0B\gg k_BT_c)$ of the Langevin model
$\mu_z=(1-kT_c/\mu_0B)\mu_0$ which allows an independent determination 
of cluster temperature and magnetic moment was never observed.

Although in many cases the Langevin model was applied very 
successfully, it was found that under certain conditions
it does not explain the experimental results correctly
\cite{Bil93}. There are two 
scenarios, where the Langevin model is likely to fail: Firstly, when the
cluster temperature is so small that only very few internal degrees of freedom
are thermally excited and therefore the heat bath  provided for 
the isothermal relaxation process is not sufficient,
and secondly, when metal clusters are investigated with an 
anisotropy energy larger than the thermal energy. In this case
the spin cannot rotate freely, but is fixed to one of the 
crystal axis (locked spin model).

Bertsch at al. proposed an adiabatic model for the magnetization of clusters \cite{Ber95}
with a magnetic
anisotropy energy   larger than the thermal energy of the
clusters (locked spin model). It is assumed that no vibrational levels
except the ground state are populated. Spin and rotation
of the clusters are treated classically. Using statistical
mechanics, the authors derived the magnetization of an ensemble of
classical symmetric rotors with the spin fixed to one of the
figure axis. In the strong field regime ($k_B T_R \ll \mu_0 B$)
 with
$T_R$ being the rotational temperature of the cluster,
the expression
\begin{equation}
\label{bersf}
\mu_z=\mu_0\left( 1-\sqrt{\frac{32}{9\pi}}
\sqrt{\frac{{k_B}T_R}{\mu_0 B}} \right),
\end{equation}
was obtained, while in the weak field regime
 the equation
\begin{equation}
\label{berlf}
\mu_z=\frac{2}{9}\frac{\mu_0 B}{{ k_B} T_R}\mu_0
\end{equation}
holds.
At large field strengths  saturation of the magnetization is predicted
(Equation \ref{bersf}). In this regime the independent determination of the rotational
 temperature and of the magnetic moment of the clusters is possible.

Up to now only one experiment, performed by Douglass et al. \cite{Blo92},
 has been discussed in the frame of the adiabatic locked spin
model. The shape of the Stern-Gerlach deflection patterns of
${\rm Gd_N}$ particles
with a bulk anisotropy energy of  $10^{6}{\rm erg/cm^3}$ 
  at 0K \cite{Gme74} generated at a nozzle temperature of $T_n=105$K 
were explained in terms of this model, but 
 no experiments on the magnetization process have been 
  performed so far. 
  In this paper we  want to investigate the  magnetization
  of clusters far away from the Langevin conditions and
   discuss the problem of isothermal versus adiabatic
  magnetization. For this purpose we study
  cold ${\rm Dy_N}$ clusters ($T_n=13$K) with a large bulk anisotropy
  energy ($10^{8}{\rm erg/cm^3}$
  at 0K \cite{Gme74}).

%\section{Experimental}
For the generation of
 ${\rm Dy_N}$  clusters (${\rm 17 \le N\le 55}$) we used a pulsed laser
 evaporation cluster source incorporated in a Stern-Gerlach molecular
 beam apparatus. The experimental setup is
described elsewhere \cite{Hih98}. The source has been modified to produce clusters 
with very small temperatures. Now it is possible to cool the nozzle 
  down to temperatures of $T_n$=13K using
 liquid He.  Additionally, the source is constructed such that 
 the dwell time of the clusters in the cold nozzle channel
 should  be sufficient  to establish  thermal equilibrium between
 clusters,  He and  nozzle.  Therefore it can be expected that 
 the cluster temperature before the adiabatic expansion equals the 
 nozzle temperature. 
 To favor a strong  adiabatic expansion through the nozzle
 into the high vacuum of the flight tubes, which  leads to further cooling of the clusters,
 a large He pressure of
about 100mbar in the nozzle channel is applied.
 In fact the velocity of the clusters
 measured behind the nozzle confirms  the  existence of 
  strong adiabatic expansion and thermal equilibrium between nozzle and clusters: 
 Using the  equation $V_{He}=\sqrt{5RT_{He}/m_{He}}$ \cite{Sco88},
where $m_{He}$ denotes the molecular weight, $R$ the
gas constant and $T_{He}$ the He temperature, we can calculate the  He terminal speed $V_{He}$, 
if we  replace $T_{He}$ by $T_n$ assuming
thermal equilibrium between nozzle and He.
 At small nozzle temperatures
 ($ 13{\rm K}\le T_n \le 40{\rm K}$)
 the calculated He terminal speed $V_{He}$ agrees very well with the 
 measured  velocity of the clusters $V_C$, while for higher temperatures 
 a growing velocity slip is observed.
For example, at $T_n$=13K we find
 $V_{C}=396$m/s$\pm 30$m/s and $V_{He}=368{\rm m/s}\approx V_C$ versus
$V_{C}=1200$m/s$\pm 40$ m/s  and $V_{He}=1765{\rm m/s}$ at $T_n$=300K. 
 The fact that  no  slip between the calculated He velocity  and the
 velocity of the clusters is observed for $T_n \le 40{\rm K}$ indicates a strong adiabatic 
 expansion and a thermal 
equilibrium between nozzle and clusters \cite{Sco88}. 
After being collimated 
the cluster beam passes the Stern-Gerlach
magnet. The deflection is detected size 
selectively by a time of
flight mass spectrometer in combination with an ionization laser
beam. The magnet and the detection unit  are  described in
\cite{Hih98}.

%\section{Results and Discussion}

\begin{figure}
\centerline{\epsfig{figure=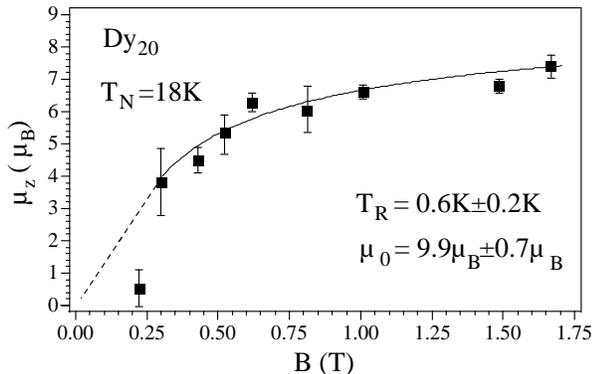, width=8cm}}
\caption{The magnetization curve of ${\rm Dy_{20}}$
  is shown. The solid line is the fitting function according to
the adiabatic model by
Bertsch et al. in the strong field limit [9]. The dotted line is the magnetization 
curve expected in the weak field limit}
\label{fig1}
\end{figure}

For ${\rm Dy_N}$ clusters generated at nozzle temperatures $T_n$=13K and 
$T_n$=18K, we observe a  shift of the Stern-Gerlach profile in the direction of 
the increasing magnetic field. This indicates that relaxation processes 
are involved. Studying the field dependence of the magnetization, 
we obtain magnetization curves which show saturation at large field strengths as 
displayed in Figure \ref{fig1}. 

To make sure that the observed saturation of the magnetization is not 
an effect of the relaxation time scales involved, we 
repeat the deflection experiments on ${\rm Dy_N}$ clusters
with a Stern-Gerlach magnet of half of the length.  We obtained
half of the deflection and therefore the same magnetization. This
shows that the measurement of the  magnetization takes place under
stationary conditions. Hence, within our resolution, there are no
relaxation processes involved with relaxation times
 in the order of magnitude of the experimental time scale of typically
200${\rm \mu}$s which is the time  ${\rm Dy_N}$ clusters need 
to traverse the Stern-Gerlach magnet.

Now we turn to the question whether we expect isothermal or adiabatic
magnetization  under the experimental conditions described above. 
To understand the nature of the relaxation process it is important
to consider which degrees of freedom are accessible in the
cluster. As pointed out above, the clusters are in thermal equilibrium with 
the nozzle before the adiabatic expansion. After the 
adiabatic expansion the clusters exhibit a
vibrational temperature $T_{vib}$ which is close to the nozzle
temperature ($T_{vib}\approx T_n\le 18$K). However, the rotational
temperature $T_R$ is   smaller, because the adiabatic cooling 
is more effective for
rotations  than for vibrations
\cite{Sco88}, since the coupling of
the rotations on the translational modes is stronger. 
 To estimate the number of vibrational states, which are thermally
 accessible in  ${\rm Dy_N}$ clusters, we calculate
  the vibrational partition sum and the occupation number of the
  vibrational ground state of the dimer ${\rm Dy_2}$.  As an  eigen frequency
we use the Einstein frequency of bulk Dy.
 We find that the ground state is occupied
with a probability of 99.9\% at $T_{vib}=18$K.  Although it is very
likely that in
${\rm Dy_N}$ clusters vibrational states with smaller
 frequencies   than the
Einstein frequency are available, this example demonstrates that
 only very few
 of the vibrational levels will be thermally accessible.
 To estimate the number of thermally accessible rotational states, 
 we calculate the
rotational occupation numbers of ${\rm Dy_{20}}$.
We approximate the complicated cluster structure by a  sphere with the
density of bulk Dy and the mass of ${\rm Dy_{20}}$.
For $T_R=1$K we obtain an  occupation maximum at
the rotational level $J_R$ =27 (2.2 \%), where $J_R$ denotes
the rotational quantum number.
Taking into account that  large quantum numbers like $J_R$=80
 are still populated (0.15\%),
 the rotation of the clusters can be described in good approximation by the model
  of the classical rotor.

In the next step we want to address the problem, whether isothermal magnetization 
is possible,  when the contribution of
the vibrations to the heat bath necessary for the 
isothermal Langevin process is  neglegible, i.e.
 when the heat bath consists exclusively of rotatational degrees of freedom.  
 This question can be answered 
 by considering the entropy transfer between
the spin and the rotational system employing the Second Law of
Thermodynamics. According to Boltzmann the entropy $S$ can be
written as $S=-\sum_i k_B \omega_i \ln \omega_i$, where
$\omega_i$ denotes the occupation probability
 of the state $i$ \cite{Lan69}.
The entropy loss $\Delta S_J$ provoked by the magnetization process
 is due to fixing the  orientation of the
total angular momentum $J$  of the clusters in the magnetic field direction z at large field 
strengths. This corresponds to the saturation of the magnetization. 
 Without field  the cluster
has $2J+1$  possibilities of equal probability $\omega_i$ to orientate the
total angular momentum versus the z-axis.  Since we observe experimentally 
the saturation of the 
magnetization (see Fig. \ref{fig1}), the order of magnitude of the magnetic moment 
$\mu_0=g_J\sqrt{J(J+1)}\mu_B$, where $g_J$ denotes the g-factor
of the cluster, can be estimated by taking into account 
that the saturation magnetization
$\mu_S$ is approximately equal to  the magnetic moment $\mu_0$.  
Assuming that  the g-factor of the clusters is in the order of
magnitude of the g-factor of the ground state of the Dy atom 
$g_j=1.33\approx g_J$, one obtains for ${\rm Dy_{20}}$ 
with $\mu_S\approx \mu_0 \approx 9 \mu_B$ a total angular momentum of 
$J\approx 6$ and hence an
entropy loss $\Delta S_J=\sum_{i=1}^{13}  \ln (1/13)k_B/13=-k_B \ln 13$. 
According to the Second Law
of Thermodynamics the rotational entropy $\Delta S_R \ge -\Delta
S_J$ has to augment at least by the same amount.  This entropy
gain changes the rotational temperature $T_R$ according to the
equation $\Delta S_R = c_{VR} \Delta T_R /T_0$, where $c_{VR}$ is
the heat capacity and $T_0$ denotes the temperature of the
clusters, before the field was applied. We have shown above that
many rotational levels with large quantum numbers  are occupied in
${\rm Dy_{20}}$. Therefore we approximate $c_{VR}$ by the high
temperature  value $c_{VR}=3k_B/2$. Hence the temperature change due to
the magnetization process is given by the expression
$\Delta T_R/T_0\ge  \ln (2J+1)2/3$.
For clusters with an initial rotational temperature $T_0\approx1$K
the temperature rises during the magnetization process  by $\Delta
T_R \approx 1.7$K. This significant  change of the rotational 
temperature shows, that in our case it is not suitable to use the isothermal 
magnetization model.

\begin{figure}
\centerline{\epsfig{figure=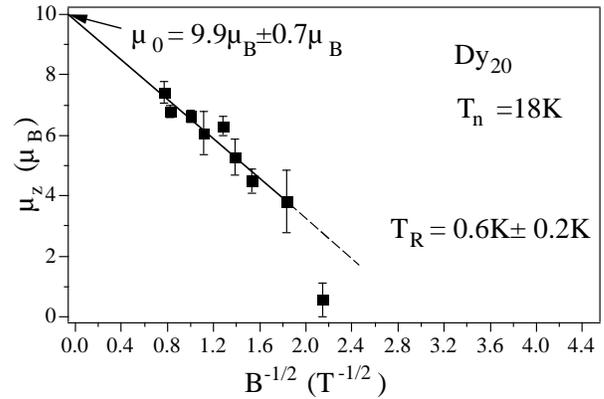, width=8cm}}
\caption{Magnetizations of ${\rm Dy_{20}}$ plotted versus
$B^{-1/2}$.
The solid
line denotes the fitting function obtained by linear regression according to
the adiabatic model. }
\label{fig2}
\end{figure}

Now let us turn to the question, whether adiabatic magnetization is possible. 
For the magnetization process the time $t_s \approx 10^{-5}$s
needed by the clusters  to move from a  zero field region into the
magnet with the full field applied, is important, because adiabatic
magnetization is only possible  when the magnetization takes place
at much larger time scales than the relaxation process. Since the
typical time scale for cluster rotations is  $10^{-9}$s, the
magnetization can take place adiabatically.
Hence, we consider whether the adiabatic model proposed by Bertsch et al. 
can be applied to evaluate our experimental data. The  assumptions made in this
model (classical rotation, classical spin and locked spin) reflect
our experimental conditions. It has been discussed above that the
rotations can be treated classically.
 In first
approximation the spin of the cluster can be treated classically as well,
 because the saturation magnetization of the clusters indicates the existence of
  large total angular momenta (see ${\rm Dy_{20}}$). The locked spin model 
  is justified, since the thermal energy of the clusters ($T_c \le 18$K) is
 much smaller than the magnetic anisotropy energy. Using the anisotropy energy 
 of bulk  rare earths ($\approx 10^8 {\rm erg/cm^3}$ at
0K \cite{Gme74}), we estimate for ${\rm Dy_{20}}$ an anisotropy energy which
corresponds to a temperature of
about 500K. 
 Therefore we use the adiabatic model proposed by Bertsch et al. to fit the 
 magnetization curves of ${\rm Dy_N}$ clusters, recorded for N=17-29
at $T_n$=18K and for N=32-55 at $T_n$=13K. 
 Since most of the
 data measured at ${ T_n=18}$K and
 at ${ T_n=13}$K belong to the strong field regime (see Fig. \ref{fig1}),
 we use Equation  \ref{bersf} to fit our magnetization curves.
 After plotting
the magnetizations versus $1/\sqrt{B}$, the magnetic moments  and 
the rotational temperatures $T_R$ 
of the ${\rm Dy_N}$ clusters are
determined by linear regression, as shown in Fig.\ref{fig2}.   In
average Equation \ref{bersf} gives a good fit for the clusters
generated at $T_n=18$K and at $T_n=13$K. Only magnetizations
measured at very small field strengths, i. e. $B$=0.22T for 
$T_n=18$K  do not fit the expected behavior, because they do not
belong to the strong field regime. $T_R$ is calculated from the
slope of the line fit and depends on the prefactors imposed by the
structural assumptions and the assumptions about the distribution
of the rotational energy.  Hence the temperature scale depends on
the geometrical details of the clusters, while the magnetic moment
is independent.

\begin{figure}
\centerline{\epsfig{figure=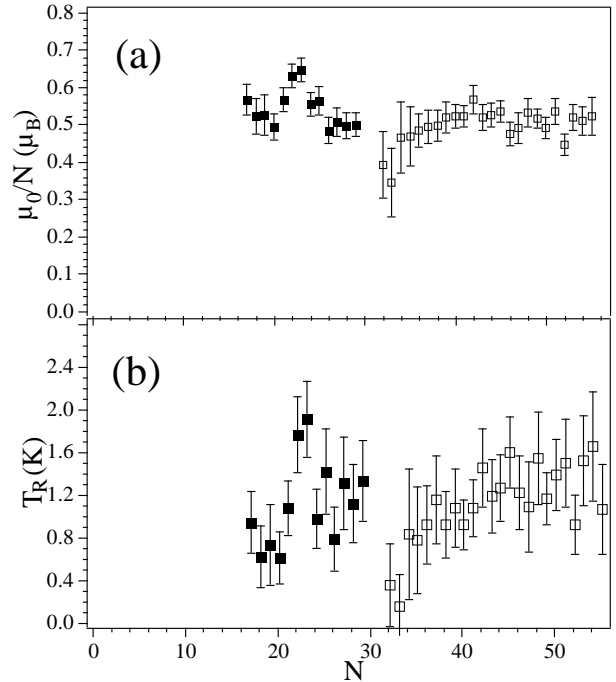, width=8cm}} \caption{In (a)
the magnetic moments per Dy atom $\mu_0/N$ of ${\rm Dy_{N}}$ clusters
  are plotted versus $N$, in (b) the
rotational temperatures $T_R$. Filled squares
denote ${\rm Dy_{N}}$ clusters generated at $T_n$=18K, empty
squares denote ${\rm Dy_{N}}$  clusters generated at $T_n$=13K.}
\label{fig3}
\end{figure}

Figure \ref{fig3}a shows the magnetic moments per Dy atom $\mu_0/N$
of ${\rm Dy_N}$ clusters with $N$=17-55 and their rotational
temperatures $T_R$ obtained by applying the adiabatic model as
described above. In average the magnetic moments per atom range
between 0.3 to 0.6 $\mu_B$. In comparison to the magnetic moment
of Dy  $\mu_{Dy}=10.6 \mu_B$ in the ferromagnetic bulk phase the
magnetic moments of the clusters are smaller by a factor of 20.
This suggests that the magnetic ordering in the clusters is rather
antiferromagnetic than ferromagnetic ($ J=\sum_i j_i\approx 0$,
with $J$ being the total angular momentum of the cluster and $j$
being the total angular momentum of the Dy atom), although the
rotational cluster temperatures are well below the Curie
temperature of the bulk ${\rm T_C}$(Dy)=86K. Keeping in mind that
the ferromagnetic ordering in rare earth metals is determined by
 indirect coupling through the valence electrons
 (RKKY interaction \cite{Kon69}), this result is not
very surprising and similar behaviour has been observed for other 
rare earth clusters \cite{Blo98}. Since the structure of small clusters
 differs strongly from
the bulk lattice to compensate surface effects, the wave functions
of non localized electrons like valence electrons change in response to the
change of the long range structural ordering. As a result the coupling between
the ${\rm 4f^9}$ cores of the Dy atoms in the clusters differs from the coupling in
the bulk, since electrons in irregularly shaped cluster orbitals are
polarized instead of electrons in orbitals described by regularly oscillating
Bloch functions. The theoretical study  performed by Pappas
 et al. on  ${\rm Gd_{13}}$ illustrates the effect of 
the modified exchange coupling on the magnetic structure 
 of small rare earth particles very well \cite{Pap96}.
In Figure \ref{fig3}b the rotational temperatures $T_R$ of the
clusters are shown.   Since the rotational temperatures (0.2-2.0K)
are much smaller than the nozzle temperatures ($T_n$=13K and 18K),
we conclude that the adiabatic cooling of the rotational degrees
of freedom after the expansion takes place very efficiently, as it
is suggested in \cite{Sco88}.

Although the adiabatic model developed by Bertsch et al. fits the
recorded magnetizations very well in the strong field limit, the
overall shape of the magnetization curve does not match the
theoretical prediction in Ref. \cite{Ber95} for a spherical
cluster. The theoretically predicted  adiabatic magnetization
curve shows a linear dependence of the magnetization at small
fields (Equation \ref{berlf}) like the isothermal magnetization, whereas the experimental
data (Fig.\ref{fig1}) suggest the existence of higher order terms.
Since it cannot be assumed that
 all ${\rm Dy_N}$ clusters in a size range of N=16-55 are spherical,
the breaking of the spherical symmetry could probably account for
the differing curve shapes at small fields, as it was shown by
quantummechanical calculations \cite{Ber96}.

%\section{Conclusion}
In summary, the magnetization of ${\rm Dy_N}$ clusters generated
at low nozzle temperatures can be understood by assuming 
adiabatic magnetization, while an isothermal process is not consistent
with the experiment.
  By measuring the saturation of the magnetization, the magnetic moment
   and the temperature of the clusters
can be determined at the same time independently. The rotational
temperatures obtained suggest a strong adiabatic cooling. The
${\rm Dy_N}$ clusters exhibit a nearly antiferromagnetic spin
order, although bulk Dy is ferromagnetic in the  temperature range
studied. This can be understood in the frame of the RKKY theory. 
The complicated size dependence of  temperature and 
magnetic moment of the clusters remains an open question.

%\section*{Acknowledgements}
S.P. gratefully  acknowledges support from the Fonds der
Chemischen Industrie. Many thanks to J.A. Becker for his 
experimental and theoretical support. 

% now the references. delete or change fake bibitem. delete next three
%   lines and directly read in your .bbl file if you use bibtex.

% \begin{figure}
% \caption{}
% \label{}
% \end{figure}


\begin{references}
\bibitem{Ste22}O. Stern
{\it Phys. Z.} {\bf13}, 3052 (1922)
\bibitem{deH90}  W.A. de Heer, P. Milani,  and  A. Chatelain,
{\it Phys. Rev. Lett.} {\bf 65}, 488 (1990)
\bibitem{Ged89}A. Gedanken, N.A. Kuebler, and M.B. Robin,
{\it J. Chem. Phys.} {\bf 90}, 3981 (1989)
\bibitem{Kue88}N.A. Kuebler, M.B. Robin, J.J. Yang, and A. Gedanken,
{\it Phys. Rev. A} {\bf 38}, 737 (1988)
\bibitem{Kan91}S.N. Khanna, and S. Linderoth
{\it Phys. Rev. Lett.} {\bf 67}, 1441 (1991)
\bibitem{Bil94}I.M.L. Billas, A. Chatelain, and  W.A. de Heer,
{\it Science} {\bf 264}, 1682 (1994)
\bibitem{Bil93}I.M.L. Billas, J. A. Becker, A. Chatelain, and  W.A. de Heer,
{\it Phys. Rev. Lett. } {\bf 71}, 4067 (1993)
\bibitem{Blo96}S.E. Apsel, J.W. Emmert, J. Deng, and L.A. Bloomfield,
{\it Phys. Rev. Lett.} {\bf 76}, 1441 (1996)
\bibitem{Ber95} G. Bertsch, N. Onishi, and K. Yabana,
{\it Z. Phys. D} {\bf 34}, 213 (1995)
\bibitem{Blo92}D.C. Douglas, J.P. Bucher, and L.A. Bloomfield,
{\it Phys. Rev. Lett.} {\bf 68}, 1774 (1992)
\bibitem{Gme74}G. Kirschstein, Gmelin Handbuch der Anorganischen Chemie,
Seltenenerdelemente, Teil B 3, Springer Verlag, Berlin, 303 (1974)
\bibitem{Hih98}T. Hihara, S. Pokrant, and J.A. Becker, 
{\it Chem. Phys.Lett.} {\bf 294}, 357 (1998)
\bibitem{Sco88}D.R. Miller in G. Scoles, Atomic and Molecular Beam Methods, 
Vol.1, Oxford, New York, 14-16  (1988)
\bibitem{Lan69}L.D. Landau, and E.M. Lifschitz, Lehrbuch der Theoretischen 
Physik, Band 1, Mechanik, Akademische Verlagsgesellschaft, Frankfurt, 134pp (1969)
\bibitem{Kon69}J. Kondo, 
{\it Solid state  physics} {\bf 23}, 184  (1969)
\bibitem{Blo98} L.A. Bloomfield, J. Deng, A.J. Cox, J.W. Emmert,
H. Zhang, D.B. Haynes, J.G. Louderback, D.C. Douglass, J.P. Bucher, and A.M. Spencer
in M. Donath, P.A. Dowben, and W. Nolting, Magnetism and Electronic Correlations
in Local-Moment Systems: Rare-Earth Elements and Compounds, World Scientific,
Singapore, 153 (1998)
\bibitem{Pap96} D.P. Pappas, A.P. Popov, A.N. Anisimov, B.V. Reddy, and S.N. Khanna,
{\it Phys. Rev. Lett.} {\bf 76}, 4332 (1996)
\bibitem{Ber96}V. Visuthikraisee, and G. Bertsch,
{\it Phys. Rev. A} {\bf 54}, 5104 (1996)
\end{references}
\end{document}